\documentclass[twocolumn,aps,prd,amsmath,amssymb,superscriptaddress]{revtex4-1}
\usepackage{graphicx}
\usepackage{multirow}
\usepackage{array}
\usepackage{dcolumn}% Align table columns on decimal point
\usepackage{bm}% bold math
\usepackage{color}

\begin{document}
\title{New Experimental Limit on the Electric Dipole Moment of the Electron in a Paramagnetic Insulator}

\author{Y. J. Kim}
\affiliation{Department of Physics, Indiana University, Bloomington IN
  47405} 
\affiliation{Physics Division, P-21, Los Alamos National Laboratory, Los Alamos, NM 87545}
\author{C.-Y. Liu}
\email{CL21@indiana.edu}
\affiliation{Department of Physics, Indiana University, Bloomington IN
  47405}
\affiliation{IU Center for Exploration of Energy and Matter,
    Bloomington IN 47408}
\author{S. K. Lamoreaux}
\affiliation{Department of Physics, Yale University, New Haven, CT 06520} 
\author{G. Visser}
\affiliation{IU Center for Exploration of Energy and Matter,
  Bloomington IN 47408} 
\author{B. Kunkler}
\affiliation{IU Center for Exploration of Energy and Matter,
  Bloomington IN 47408}
\author{A. N. Matlashov}
\affiliation{Physics Division, P-21, Los Alamos National Laboratory, Los Alamos NM 87545}
\author{J. C. Long}
\affiliation{Department of Physics, Indiana University, Bloomington IN
  47405}
\affiliation{IU Center for Exploration of Energy and Matter,
    Bloomington IN 47408}
\author{T. G. Reddy}
\affiliation{IU Center for Exploration of Energy and Matter,
  Bloomington IN 47408} 
\affiliation{Department of Pure and Applied Physics, Guru Ghasidas University, Bilaspur, India}

\date{\today}
\begin{abstract}
We report results of an experimental search for the intrinsic Electric
Dipole Moment (EDM) of the electron using a solid-state technique. The
experiment employs a paramagnetic, insulating gadolinium gallium
garnet (GGG) that has a large magnetic response at low
temperatures. The presence of the eEDM would lead to a small but
non-zero magnetization as the GGG sample is subject to a strong
electric field. We search for the resulting Stark-induced
magnetization with a sensitive magnetometer.  
Recent progress on the suppression of several sources of background
allows the experiment to run free of spurious signals at the level of
the statistical uncertainties. We report our first limit on the eEDM
of $(-5.57 \pm 7.98 \pm 0.12)\times$10$^{-25}$e$\cdot$cm with 5 days
of data averaging. 

\end{abstract}

% insert suggested PACS numbers in braces on next line
%\pacs{04.50kd}
% insert suggested keywords - APS authors don't need to do this
%\keywords{}

\maketitle

\section{Introduction}
The search for the electric dipole moment (EDM) of elementary
particles is motivated to test the discrete symmetries assumed in the
Standard Model (SM) of particle physics.  
On account of the different transformation properties of the EDM (a
polar vector) and the spin (a pseudo-vector), the fundamental physical
laws governing particles must violate both time-reversal (T) and
parity-inversion (P) symmetries for a fermion to acquire an
EDM~\cite{Purcell50}.  
While the phenomenon of P violation is firmly established in numerous
experiments, T violation has only been observed directly in the
neutral kaon~\cite{CPlear98} and neutral $B$ meson~\cite{Babar12} systems.  
Measurements of EDMs of elementary particles use different
experimental techniques, often on low energy systems at $Q\simeq 0$,
to probe the physics of T violation (with no flavor-changing) at
energy scales higher than tens of TeV, and could provide information
complementary to that from high-energy collider experiments on the
nature of symmetry breaking.  

The physics of T violation is often linked, via the CPT theorem, to
the violation of the combined Charge conjugate (C) and P symmetry. 
The only confirmed source of CP violation in the SM is the complex
phase of the CKM matrix (that describes quark mixing in
charged-current weak interactions). With it,  
the electron EDM (eEDM) manifests through high-order loop couplings
that involve flavor-changing quark interactions with the exchange of
W$^{\pm}$ weak bosons.  
The resulting size of the eEDM predicted within the framework of the SM is no
larger than $10^{-38}$~e$\cdot$cm,    
which is well beyond the reach of current experimental techniques. The
current experimental upper bound is established by measuring electron spin
precession in pulses of thorium monoxide molecules, with a sensitivity of
8.7$\times$10$^{-29}$e$\cdot$cm~\cite{Acme14}.  
New sources of CP violation introduced by theories beyond the SM often
lead to a sizable eEDM.  
Free from the SM backgrounds, the measurement of EDM presents a
powerful tool for global tests of many theoretical extensions to the
SM.  
In particular, some variants of the popular supersymmetric model can
generate EDMs of elementary particles comparable to the current
experimental limit, and will be put to stringent tests as the next
generation of experiments improves the sensitivity by another factor
of 100.  
Even though none of the experimental efforts have yielded positive
results, the EDM searches, to this end, have ruled out many
theoretical models. With ever-more-refined experimental techniques,
the EDM searches continue to be of fundamental significance in
particle and nuclear physics. 

The conventional technique used to measure EDM is the 
separated oscillatory fields method of Ramsey~\cite{Ramsey50}
based on nuclear magnetic resonance: the EDM interactions induce an
additional frequency shift in the Larmor precession, when the particle
under study is subject to an electric field applied parallel (or
anti-parallel) to a weak magnetic field.  
In the attempt to improve the experimental limit on the eEDM, we have
been pursuing an alternative approach using a solid-state
technique~\cite{Shapiro68}.  
The application of a strong electric field to a paramagnetic insulator
would align the EDMs of valence electrons bound in the solid, leading to a
small yet non-zero magnetization. Even though the energy shift
predicted from the EDM coupling for individual electrons is much
smaller than the thermal energy, the cumulative effect from the large
number of electrons in a solid sample leads to a net spin alignment
equivalent of a few million Bohr magnetons. This Stark-induced
magnetization can be detected using sensitive magnetometry.  A result
using this method was recently realized using the paramagnetic
ferroelectric Eu$_{0.5}$Ba$_{0.5}$TiO$_{3}$, with a sensitivity of
$d_{e}<6.05\times$10$^{-25}$e$\cdot$cm~\cite{Eckel12}.  The original
material proposed for this technique was the paramagnetic insulator
Gadolinium Gallium Garnet (GGG, Gd$_3$Ga$_5$O$_{12}$). In this paper,
we report a measurement of $d_{e}$ in GGG with a sensitivity within a
factor of 2 of~\cite{Eckel12}.

The use of Gadolinium Gallium Garnet (GGG, Gd$_3$Ga$_5$O$_{12}$) for
an eEDM search was first proposed in~\cite{Lamoreaux02}; experimental
construction was reported in~\cite{Liu04}.   
The GGG material is advantageous for the high number density of the
Gd$^{3+}$ ions ($\sim$10$^{22}$cm$^{-3}$), each containing seven
unpaired electrons in the $4f$ shell, leading to a strong magnetic
response in a bulk sample. In addition, the GGG possesses superb
dielectric strength of 10~MV/cm and a high electrical resistivity that
allows it to withstand strong electric fields (necessary to perform
eEDM measurements) with sufficiently small leakage currents.  

Ideally, the experiment has to be performed in an environment free of
magnetic fields, because the spin coupling to the magnetic field would
certainly dominate over the small EDM interaction. 
In practice, even with the most hermetic magnetic shielding, some
residual field is inevitable. Therefore, the experiment is carried out
in AC mode, in which the change of magnetization upon the reversal of
the electric field is measured.   
Unfortunately, during the field reversal, transient currents create a
magnetic field that also flips direction. The transient field can die
down quickly, but 
the presence of hysteresis effects that lead to finite remnant
magnetization (with long relaxation time) in the sample would be
detrimental to the successful realization of this technique at the
proposed sensitivity level. 
Therefore, we use a paramagnetic garnet as a precaution against
possible systematic effects that would likely arise with the use of
ferromagnetic materials, even though the magnetic susceptibility is
much higher in iron garnets.

\section{Sample Characterization}
The garnet family~\cite{Geller67} has a general
structure $\{A_3\}[B_2](C_3)O_{12}$, where $A$ denotes
triply-ionized metallic ions, M$^{3+}$, on a $\{$dodecahedral$\}$
lattice, and $B$ and $C$ are ions on $[$octahedral$]$ and
$($tetrahedral$)$ lattices, respectively.  
Oxygen ions, O$^{2-}$, form a cage around the ions and balance the charge.
In general, the couplings between these sub-lattices are
anti-ferromagnetic (AF). 
The $A$, $B$, and $C$ ions can be substituted by many different metallic
elements, with varying degrees of magnetism, leading to a wealth of
magnetic properties that change with temperature. The rare-earth
elements, which are of interest to the eEDM search, can occupy the $A$
sites. Due to relativistic effects, the EDMs of paramagnetic atoms
(ions) is enhanced by a factor of $Z^3$, where $Z$ is the atomic number.   
Thus, the EDM of the Gd$^{3+}$ ion ($Z=64$) dominates over that of the
Ga$^{3+}$ and O$^{2-}$. Non-magnetic Ga$^{3+}$ ions populate the $B$
and $C$ sites, leaving the magnetic property of GGG to be determined
solely by the spin interactions of the Gd$^{3+}$ ions on the
dodecahedral lattice.    
Despite the intrinsic AF coupling, the spin of the Gd$^{3+}$ remains
disordered and follows a typical paramagnetic behavior with which the
magnetic susceptibility increases with decreasing temperatures. Only
at temperatures lower than a few hundred milli-Kelvin does the AF
coupling prevail, and the system becomes geometrically frustrated,
transforming to a spin glass state~\cite{Schiffer95}. However, the
spins are never fully ordered and the spin degree of freedom remains
unfrozen at low temperatures~\cite{Dunsiger00, Marshall02}.  

\begin{figure}[t]
\centering
\includegraphics[width=3.4in, height=2.5in]{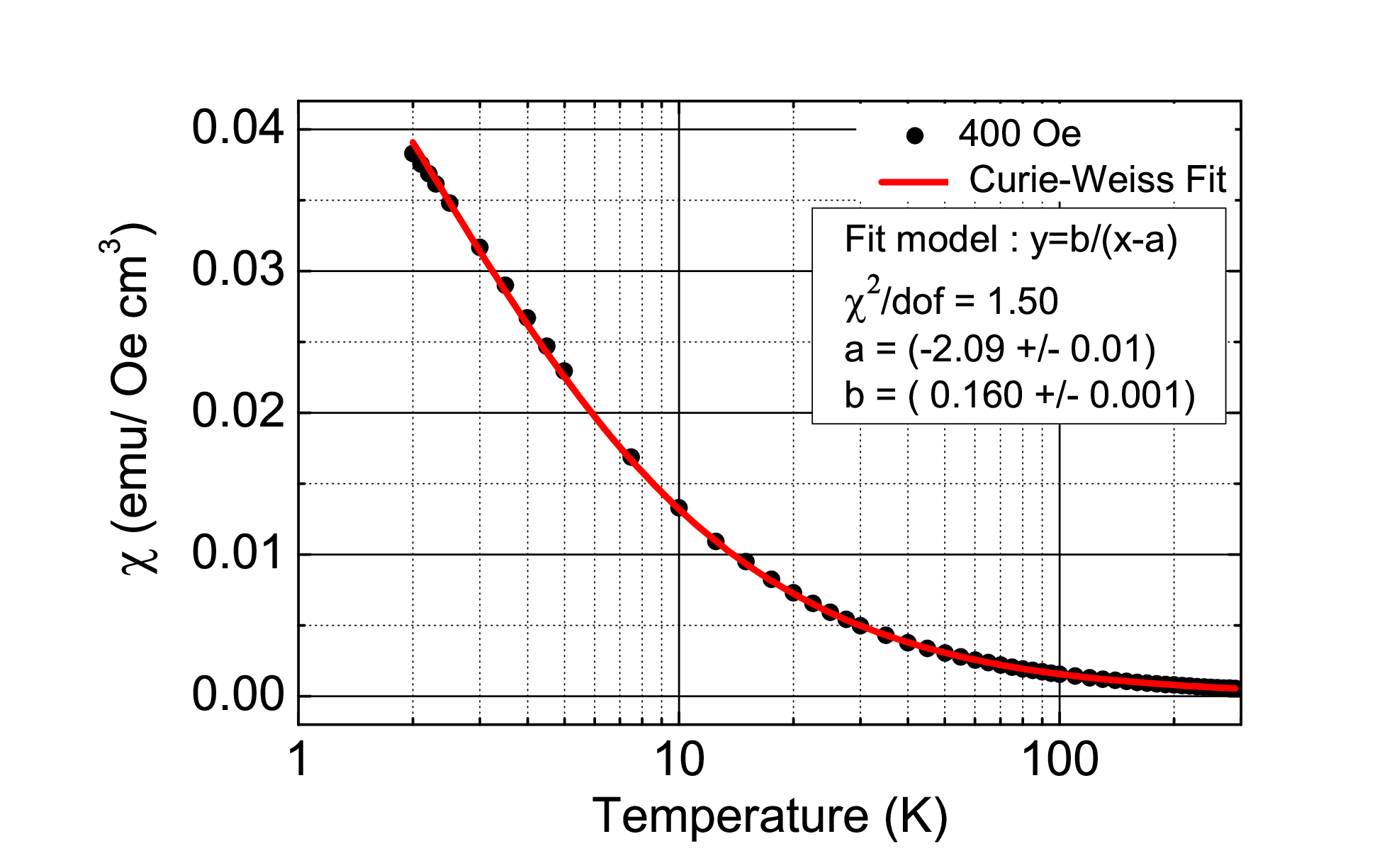}
\caption{\label{1}Volume magnetic susceptibility of the GGG sample as a function of temperature, measured at a maximum applied magnetic field of 400~Oe. The solid curve shows the Curie-Weiss fit.}
\end{figure}

We measure the magnetic susceptibility $\chi$ of polycrystalline GGG
samples (synthesized in our lab~\cite{Liu04} using solid-state
reactions~\cite{Hellstrom89}) using a SQUID susceptometer
system\footnote{Quantum Design Magnetic Property Measurement System
  (MPMS)}.   
The volume magnetic susceptibility was measured at several values of
the applied magnetic field (400~Oe, 100~Oe, and 10~Oe).  
The result with the maximum applied field of 400~Oe is shown in
Fig.~\ref{1} from 295~K to 2~K.  
The fit to the Curie-Weiss relation, $\chi= C/(T-\theta_{CW})$, gives
a Curie-Weiss temperature of -2.1~K, indicating that coupling of
adjacent Gd$^{3+}$ ions is indeed AF .  
The strong AF coupling could lead to an order-disorder phase
transition at low temperatures, and limit the size of
$\chi$. Fortunately, for GGG, this phase transition is highly
suppressed and was never observed due to the geometric frustration of
AF-coupled spins on a Kagome lattice~\cite{Schiffer95}. To maintain a
high sensitivity to the eEDM, the spins need to remain free to respond
to the external fields, therefore, it is essential to learn more about
the conditions of phase transitions to ensure that the experiment is
operated with the GGG in the paramagnetic phase.  

Finally, to assess $\chi$ of the material, a correction for the
demagnetization effect on the measured susceptibility $\chi_{mea}$ is
applied:  
\begin{equation}\label{chi}
\chi = \frac{\chi_{mea}}{1-N\chi_{mea}}.
\end{equation}
Here $N$ is the demagnetizing factor, which arises from 
an additional demagnetization field created by the magnetic surface
charge density $\sigma_M=\hat{\mathbf{n}}\cdot\mathbf{M}$. This leads
to a partial cancellation of the applied field inside the sample,
analogous to the electrostatic depolarization field.  
Our cube-shaped sample with dimensions of
0.3~cm$\times$0.3~cm$\times$0.3~cm is estimated to have $N=0.264$,
using finite-element-analysis calculations. 

We also measured the resistivity of the synthesized polycrystalline
GGG sample using an electrometer (Keithley 6517B), and observed a
volume resistivity of $(5.32\pm0.04)\times10^{15}$ $\Omega\cdot$cm and
a surface resistivity of $(2.95\pm0.02)\times10^{15}$ $\Omega$/cm$^2$
at 300~K. 

\section{Experimental details}
\begin{figure}[t]
\centering
\includegraphics[width=3.5in]{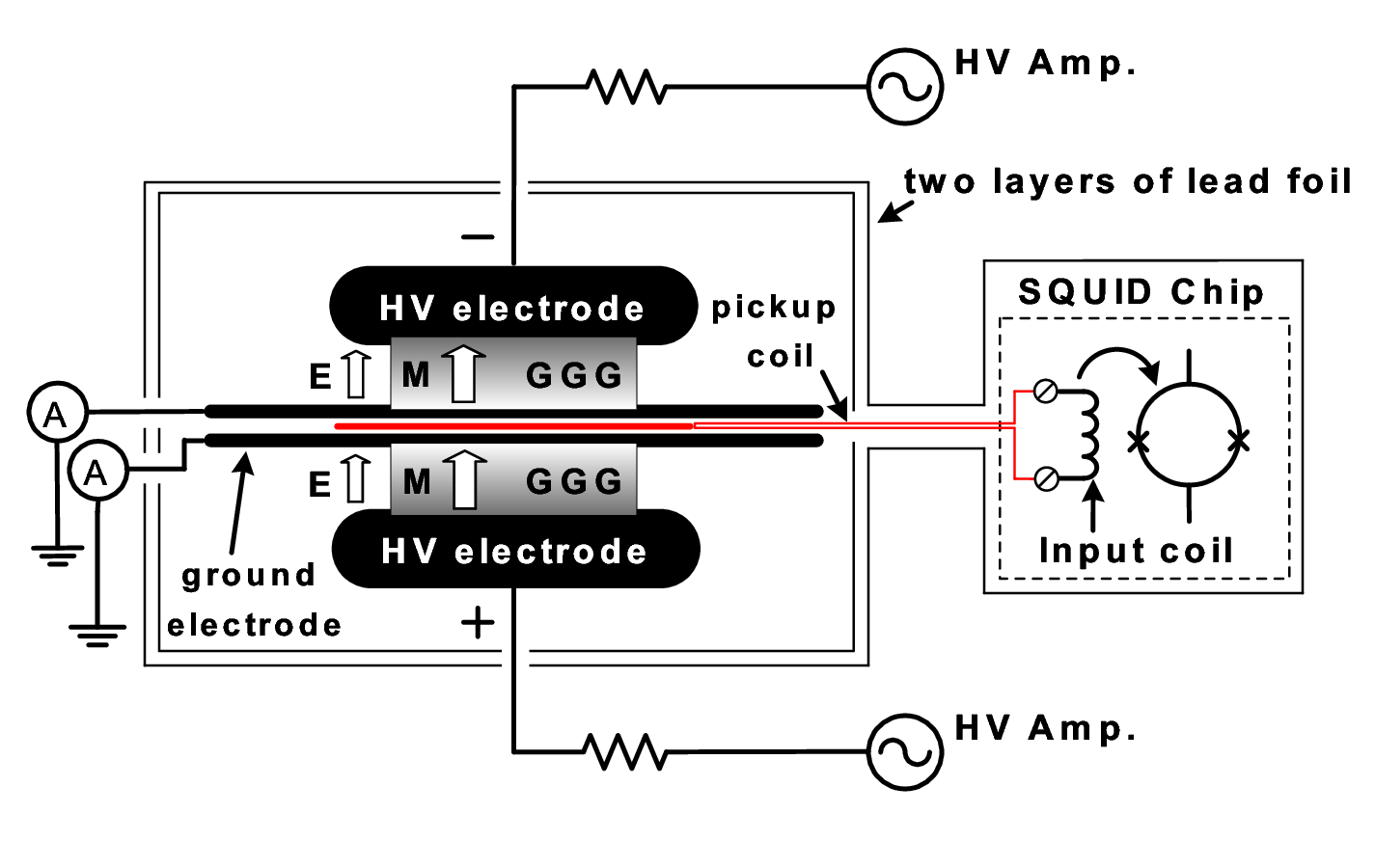}
\caption{\label{setup}Schematic of the experimental setup for the GGG-based solid-state eEDM experiment}
\end{figure}

With the samples characterized, we built the
EDM experimental cell from two disk-shaped GGG samples, each with
diameter 3.3~cm, thickness 0.76~cm, and density 6.66~g/cm$^3$, sandwiched
between two planar electrodes and two isolated ground plates
(Fig.~\ref{setup}). 
The electrodes are connected to high voltage (HV) sources of opposite
polarities in such a way that the electric fields in both GGG samples
are parallel. 
In the presence of a strong electric field, the eEDMs are aligned by
the electric field, leading to a net spin-polarization, because the
EDM vector is bound in the same (or opposite) direction of the spin
vector as a result of the Wigner-Eckart theorem. 
 This Stark-induced spin ordering generates a bulk magnetization which
 produces a magnetic field surrounding the paramagnetic GGG sample.  
The magnetization can be detectable using a DC superconducting quantum
interference device (SQUID), serving as a flux-to-voltage transducer.  

Due to the small energy of eEDM interactions, the induced magnetic
field is minute. However, with EDMs near the current experimental
limit, the accumulated magnetic signal from the large number of
electrons inside the GGG sample could lead to a sizable signal above
the background.  
To be more explicit, with an internal field $E_{int}$ of 10~kV/cm, the
EDM interaction energy, $d_eE_{int}\simeq 10^{-24}$~eV, using the
current upper limit for the eEDM, $d_e$.  
In contrast, the thermal energy is as large as $k_BT =
8.6\times10^{-7}$~eV at 10~mK, where $k_B$ is the Boltzmann constant.  
The degree of EDM-induced spin alignment is washed out by the thermal
fluctuation to the level of $d_eE_{int}/{k_BT}\simeq 10^{-17}$ for
each Gd$^{3+}$ ion. 
On the other hand, the large number of Gd$^{3+}$ ions (each with a
magnetic moment $\mu_a$) in the solid results in a bulk magnetization
of 10$^{22}$cm$^{-3}\times$10$^{-17}\mu_a \simeq 10^{5} \mu_B$,
leading to a net magnetic field of 10$^{-18}$~T.  

\begin{figure}[t]
\centering
\includegraphics[width=2.5in]{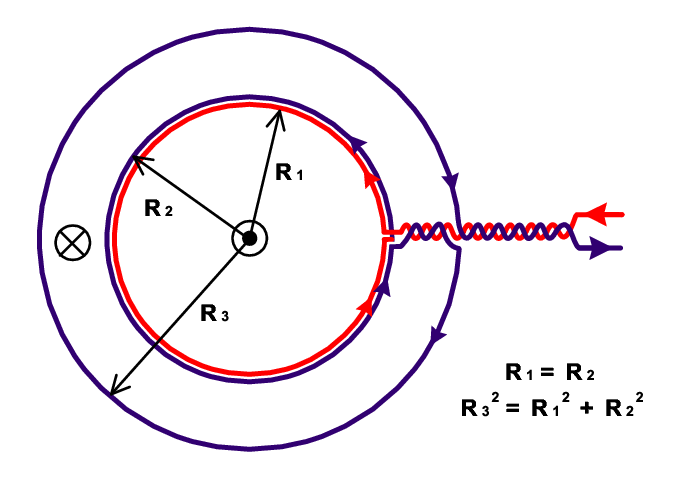}
\caption{\label{pickup} The magnetic flux pickup coil in the form of a planar gradiometer. It is placed between the ground electrodes. The inner diameter matches the diameter of the GGG sample. For best CMMR, the inner area $\pi R_1^2$ is matched to the outer area $\pi R_3^2-\pi R_2^2$.}
\end{figure}

A flux pickup coil, in the form of a planar gradiometer, is sandwiched
in between the ground planes. It integrates the EDM-induced magnetic
flux over the area of the coil. More importantly, the use of a
gradiometer eliminates the common-mode magnetic signal from the
residual magnetic field remaining inside the magnetic shield. It can
significantly reduce the magnetic pickup due to the vibrational
motions of the coil in a residual field.  
The gradiometer (Fig.~\ref{pickup}), with a two-turn inner coil wound
clockwise and a single-turn outer coil wound counter-clockwise, is
optimized to have the proper inductance matching to the input coil of
the SQUID sensor. The diameter of the inner coil matches the diameter
of the sample.  
The common-mode rejection ratio (CMMR) of a typical hand-wound coil
was measured to be $\sim$ 200, corresponding to a 0.5\% area
mismatch. 
The magnetic flux pickup is slightly enhanced by partially enclosing
the returning flux, leading to an enhancement factor of 1.1
(calculated using finite-element analysis) compared to that using a
simple one-turn coil.  
In other words, the effective area $A$ for flux pickup in this
gradiometer is a factor of 1.1 higher than the actual cross-sectional
area of the sample.  
This enhancement factor can be increased up to 1.8 by reducing the
radial dimension of the superconducting lead shield, and thus
compressing the return flux lines laterally to increase the flux
pickup.  

The EDM-induced magnetic flux in the sample enclosed by the pickup
coil, $\Phi_{e}$, can be estimated using~\cite{Lamoreaux02} 
\begin{equation}\label{phi}
\Phi_{e} = f\cdot \frac{\chi \alpha d_e E_{ext} }{\mu_a}\cdot A.
\end{equation}
Here $\alpha$ is the paramagnetic EDM enhancement factor for the Gd$^{3+}$
ion in the GGG structure, which includes the effect of the dielectric
reduction of the external field. $E_{ext}$ is the strength of the
externally applied electric field,  
$A$ is the effective area of the pickup coil, $\mu_a$
is the magnetic moment of the paramagnetic Gd$^{3+}$ ion
($\mu_a=g\sqrt{J(J+1)}\mu_B =7.94 \mu_B$, with $g=2$ and $J=7/2$), and
$f$ is the flux suppression factor including the demagnetizing effects
depending on the geometry of the GGG sample. 

The effective EDM enhancement factor $\alpha$ in Eq.~\ref{phi} was
calculated by Sushkov's group~\cite{Buhmann02, Dzuba02, Kuenzi02,
  Mukha03}.  
In the perturbative calculation, 
the EDM-induced energy shift $\Delta\epsilon$ per Gd$^{3+}$ ion arises
from three independent effects, including   
(a) the EDM enhancement in the Gd$^{3+}$ ion, (b) the
electron-electron coulomb interaction, and (c) the lattice deformation
and the positional shift of the Gd$^{3+}$ ion with respect to the
surrounding O$^{2-}$ ions in the GdO$_8$ cluster.  
The resulting $\Delta\epsilon$ is $35.6d_eE_{int}$ where $E_{int}$ is
the internal electric field inside the sample. On account of the
dielectric reduction of the internal electric field
($E_{int}=E_{ext}/K$, where the dielectric constant $K\approx12$ in
GGG~\cite{Lal77,Shannon90}), the overall energy shift per Gd$^{3+}$ ion
is  
\begin{equation}\label{shift}
\Delta\epsilon = 35.6d_e\left(\frac{E_{ext}}{12}\right)=2.97d_eE_{ext}.
\end{equation}
Note that the authors used a value of $K=30$, quoting from an online table
~\footnote{\url{http://www.mt-berlin.com/charts/chart_07.htm\#IND9}. The
  same company also has another table:
  \url{http://www.mt-berlin.com/frames_cryst/descriptions/substrates.htm},
  containing a different value of K = 12 for GGG.}.  
The references we found all suggest a smaller $K\simeq 12$, and we
also independently confirmed the smaller dielectric constant of the
GGG with capacitance measurements.  
Hence, the effective EDM enhancement factor $\alpha$ is 2.97. This is
a factor of 3.5 larger than the original estimate, in which the energy
shift is:
\begin{eqnarray}
\Delta\epsilon & = &
               -d_{a}E_{l}=-(-2.2d_e)\frac{1}{3}(K+2)(E_{ext}/K)\nonumber
               \\
               & = & 0.86d_eE_{ext},\nonumber 
\end{eqnarray}
where $d_{a}=-2.2d_e$~\cite{Dzuba02} and $E_{l}$ is
a local field acting on Gd$^{3+}$, estimated using a simple Lorentz
relation $E_{l}=\frac{1}{3}(K+2)E_{int}$.  

The flux suppression factor $f$ in Eq.~\ref{phi} describes the degree
to which the actual flux measurable from the EDM-induced magnetization
is reduced due to the effect of geometry.  
The suppression factor $f$ of the disk-shaped GGG sample with a 3.3~cm
diameter and a 0.76~cm height is calculated using finite-element
analysis. The solution shows a non-uniform magnetic field given a
uniform magnetization inside the disk-shaped sample. To obtain
$\Phi_e$, we integrate the solution field over the area of the pickup
coil.  
The resulting magnetic flux is suppressed by a factor of 0.369 due the
finite dimensions of our sample geometry, and by 
another factor of 0.425 due to the placement of the pickup coil
0.33~cm away from the surface of the sample. Note that this
latter reduction is of a different origin from the demagnetization
effect discussed before in the context of the magnetic susceptibility
measurement.   
The total suppression factor $f$ is estimated to be 0.157, leading to
a loss in sensitivity to the eEDM not considered in the original
proposal~\cite{Lamoreaux02}. Note that this suppression factor can be
improved by  moving the pickup coil closer to the sample. 

As shown in Fig.~\ref{setup}, the pickup coil (with an inductance
$L_p$ of 618~nH) connects to the built-in input coil on the SQUID
sensor chip (Superacon CE2blue, with an input coil inductance $L_i$ of
420~nH).  The mutual inductance $M$ between the input coil and the
SQUID is 8.1~nH.  
The increase in magnetic flux $\Phi_{e}$ picked up from the sample
induces a current $I=\Phi_{e}/(L_p+L_i)$. The current flows into the
input coil, and produces a flux $\Phi_{sq}$ that couples into the
SQUID loop inductively, and is read out as a voltage signal. The
relationship between $\Phi_{sq}$ and $\Phi_{e}$ is given by 
\begin{equation}\label{squid}
\Phi_{sq} = MI= \frac{M}{L_p+L_i}\Phi_{e}=\beta\Phi_e,
\end{equation}
where $\beta$ is the coupling efficiency which quantifies how much the
flux is diminished when $\Phi_{e}$ is delivered to the SQUID
sensor. The coupling efficiency is calculated to be 0.0078 in our
setup.  
To enhance the measurable flux, we need a strong electric field,
a large sample size, and an optimized pickup coil (see
Eq.~\ref{phi}). 

\begin{figure}[t]
\centering
\includegraphics[width=3.2in, height=2.3in]{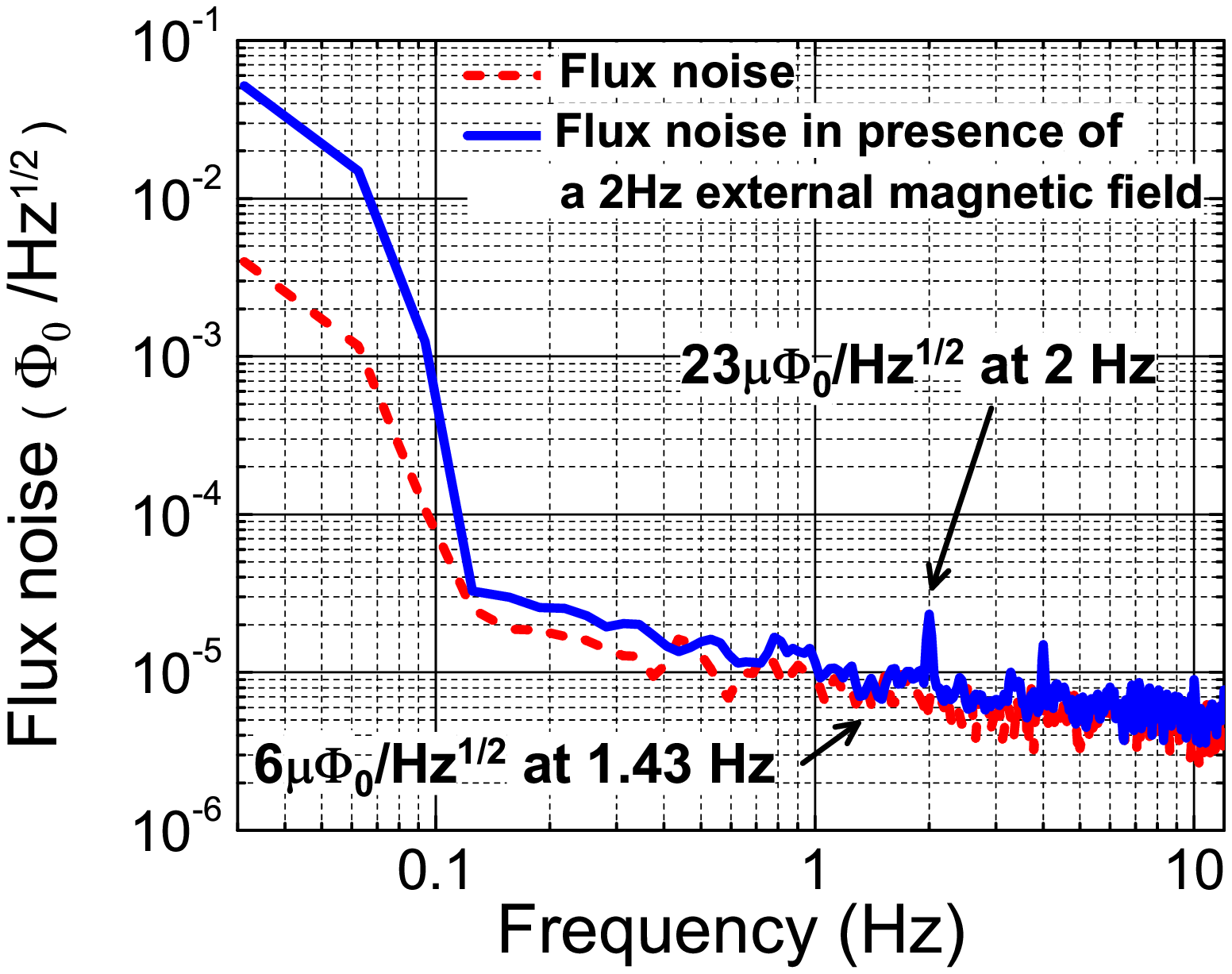}
\caption{\label{fig:superacon} Magnetic noise spectra of the DC SQUID
  detector operated at 4.2~K, showing nominal noise spectrum (dashed
  curve), and measured spectrum with an external magnetic field of
  10~gauss at 2~Hz, applied outside the magnetic shield.} 
\end{figure}

To reduce Johnson noise,  
the electrodes are made from machinable ceramics (MACOR) coated with
graphite to provide large but finite resistivity. The large electrical
resistivity helps to reduce eddy currents and the magnetic noise
produced by random motion of conducting electrons.  
During the eEDM measurement, voltages of opposite polarities are applied to
the two HV electrodes so that the electric fields in both samples are
in the same direction.  
The leakage current on each ground plate is monitored by 
a dedicated low-noise current preamplifier (Stanford Research Systems SR570).
The assembly of the GGG samples and electrodes is shielded from
external magnetic fields with two layers of superconducting lead foils
and three additional layers of mu-metal (Metglas alloy ribbon) wound
on square forms with symmetry axes along the $x$, $y$, and $z$
directions, respectively. The experiment is mounted inside a helium
cryostat to allow full immersion in a bath of liquid helium at 4.2~K
at atmospheric pressure: all eEDM measurements are taken at
4.2~K. Finally, a cylinder of Co-Netic ferromagnetic 
shielding at room temperature surrounds the whole cryostat to
provide initial reduction of ambient fields. 
The noise spectrum of the SQUID sensor as instrumented inside the working
experiment is shown in Fig.~\ref{fig:superacon}. 
Note that the experimental setup described here is the prototype
design for the proof-of-principle measurement. 
Future work is planned to extend this work to sub-Kelvin temperatures,
in order to attain better sensitivity through the enhanced magnetic
susceptibility. 

\section{Essential Improvements}
\label{sec:improvements}
For stable operation of the SQUID sensor at the base noise level, any
uncontrolled sources of electromagnetic interference (EMI) are
undesirable. Considerable effort went into studying and eliminating
electrical EMI in the lab, and eliminating ground loops. As
shown in Fig.~\ref{fig:superacon} (dashed curve), the baseline of the
SQUID sensor is $6\mu\Phi_0/\sqrt{Hz}$ at the frequency of operation
(close to the manufacturer's specification).  The solid curve
shows the spectrum in the presence of a 2~Hz external uniform
magnetic field of 10~gauss, generated by a pair of Helmholtz coils
placed outside the cryostat. By comparing the amplitude of the
residual 2~Hz peak to the applied field strength, we estimate the
overall magnetic shielding factor of the system to be
$5\times10^{11}$. 
The quality of the magnetic shield and the ground-loop isolation also
ensures that no observable vibrational peaks or power line harmonics
show up on the SQUID noise spectrum. 

The expected EDM-induced magnetic signal to be measured by the SQUID
sensor is small, thus any possible contamination from other voltage
monitoring channels (such as the high voltage channel, which has a
very large signal by comparison) through capacitive coupling is
intolerable. 
To address this problem, a custom data acquisition (DAQ) system was
built, with the capability of eight-channel simultaneous
sampling~\cite{Kim2011a}.  
The DAQ system has eight dedicated analog-to-digital converters (ADC)
with 24-bit resolution for each analog input channel, individually
shielded in its own isolated heavy-duty radio frequency (RF) shielding
enclosure. 
Fiber optic links are implemented between each satellite ADC board and
the master board for measurement control and data retrieval.  
Using this DAQ system, we are able to minimize cross-talk between
channels to better than -191~dB.  
The system is carefully designed to reduce EMI, and eliminate the
possibility of unwanted currents 
flowing in ground loops. 
The intrinsic root-mean-square noise of the DAQ system is measured to
be 1.38~$\mu$V. The DAQ system allows us to collect a large amount of
data for averaging,  
without introducing additional sources of non-Gaussian noise at the
level of the desired voltage sensitivity. The Gaussian-distributed
random noise from the SQUID detector can thus can be reduced with
higher statistics. 

\begin{figure}[t]
\centering
\includegraphics[width=3.2in, height=2.3in]{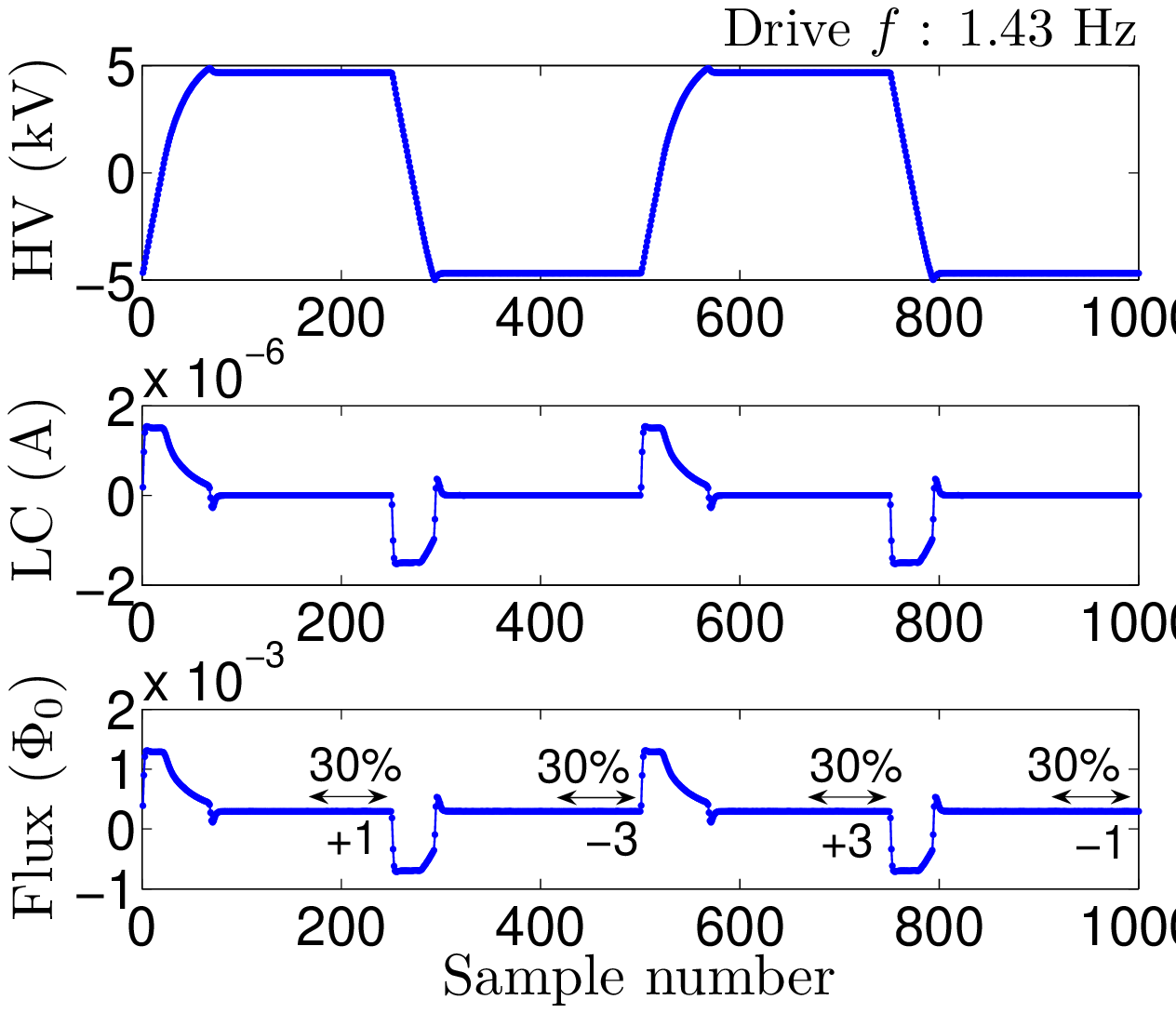}
\caption{\label{signal} Time traces of averaged signals on one HV monitor, one leakage current monitor, and the SQUID readout (over three hours of sampling) presented as two cycles of field reversal. Data were taken using the 24-bit DAQ system, with a sampling rate of 715~Hz.}
\end{figure}

This low-noise, high-resolution DAQ system enables us to  
study the correlations between the measured leakage current and
applied high voltage.  We usd this system to uncover a systematic
effect originating from 
the voltage drift of the HV polarity switch.  
After each polarity reversal, the HV settling time could be quite long.
With a small voltage drift on top of the nominal 5~kV charging
voltage, the unquenched charging and discharging of the electrodes
leads to non-zero electric currents $C dV/dt$ flowing in and out of
the electrodes. This current is detected by the leakage current
monitor. It also generates a magnetic field in phase with the polarity
of the high voltage, given that the normal of the
pick-up coil is not perfectly aligned with the field lines. This leads
to a signal (on the SQUID channel) that mimics the EDM-induced magnetization.
To eliminate this source of background, we improved our HV polarity
switch system to have distortion $<$10~ppm/sec of voltage drift.  
In order to handle the polarity switching between $\pm$5 kV, we use
vacuum-tube triodes (6BK4C) connected in series to the positive and
negative DC HV supplies (Stanford Research PS350). The gate voltage
for each vacuum tube is controlled through a opto-isolator, driven by
an arbitrary waveform generator; the input square waveform can be
amplified by the HV polarity switch system by a factor of 1000. 
%{\color{red} It consists of each positive and negative DC power supply (Stanford Research PS350), four HV vacuum tube triodes (6BK4C) for the function of HV polarity switch, a function generator for the voltage drive, and feedback circuits.} 
A reduction of the voltage drift by a factor of 600 over the previous
supply was achieved by improving the feedback circuit and reducing the
transient ramp at field reversals.  
Currently, the voltage drift on the HV output is limited by the drift
of the low-voltage drive from the arbitrary function generator, which
has only 14-bit resolution. Our next-generation HV system uses a
precision 22-bit DAC drive (integrated into the DAQ system) and should
improve the $dV/dt$ by another factor of 30, with increased HV output
from 10~kV$_{pp}$ to 40~kV$_{pp}$. 

In addition, early problems we encountered with SQUID instabilities,
which resulted in frequent flux jumps at applied electric fields higher
than 3~kV/cm, were resolved by replacing the HV cables and adding low-pass
filter resistors, which needed to be cryogenically-compatible and
HV-rated.  
This improvement allows the application of the full range of
10~kV$_{pp}$ across the sample.  

\section{Analysis}
\begin{figure}[t]
\centering
\includegraphics[width=3in, height=2.3in]{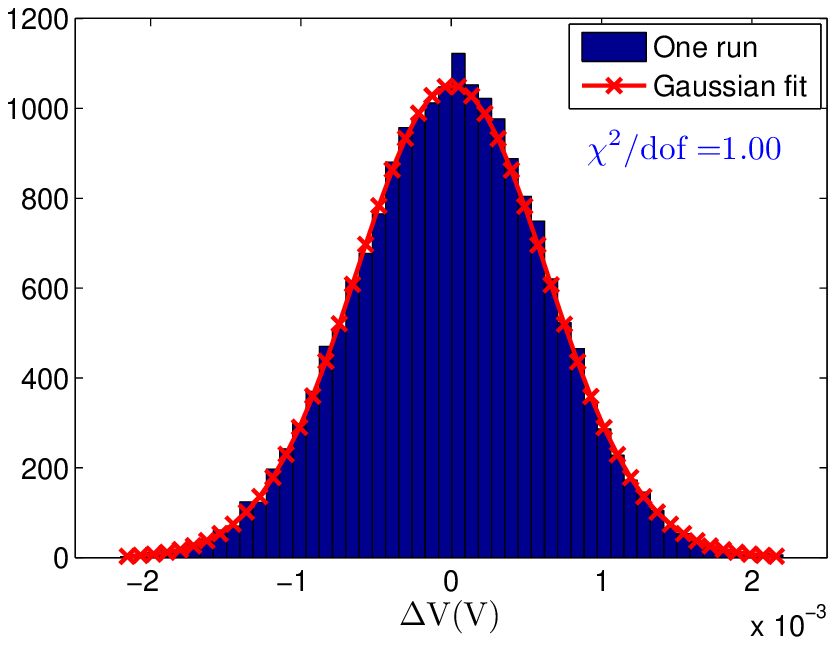}
\caption{\label{fig:histogram}Distribution of the eEDM observable (processed with the drift-correction algorithm). Data points were collected over one non-interrupted run (3 hours).}
\end{figure}
 
Typical averaged time-traces of three hours of data on the monitored
channels are plotted in Fig.~\ref{signal}. In an eEDM measurement
sequence, we apply a voltage of alternating polarities up to
$\sim$10~kV$_{pp}$ in a square waveform with controlled ramp speed on
the electrodes. The polarity switching cycle is repeated at a rate of
1.43~Hz. This drive frequency is chosen to be low enough to reduce the
transient currents, but high enough to avoid the low frequency 1/$f$
corner of the SQUID noise spectrum (Fig.~\ref{fig:superacon}). Using
the 24-bit DAQ system, we monitor the high voltage on the two
electrodes (through 1000:1 voltage dividers), currents flowing in the
two isolated ground plates, and the analog voltage output from the
SQUID readout electronics that could be converted into the magnetic
flux $\Phi_{sq}$ through a predetermined transfer function.  

The time-trace of the current monitor contains both the charging and
discharging current ($CdV/dt$) and the leakage current ($V/R$) flowing
through the bulk sample or the surface.  As shown in Fig.~\ref{signal},
the measured currents through the ground plates are dominated by the
charging/discharging transient currents during the HV polarity
switching. The SQUID sensor measures magnetic fields, generated by the
EDM-induced sample magnetization as well as by the electric currents
flowing in and out of the electrodes. During field reversals, the
SQUID is measuring the large magnetic fields associated with transient
currents: these must dissipate sufficiently quickly for the SQUID to
discern the EDM-induced magnetic flux once the field settles to the maximum
amplitude of the applied HV.  
The difference of the magnetic fluxes at +HV and -HV within one cycle
is proportional to the eEDM, and is defined as our eEDM observable.  
The $dV/dt$ needs to be controlled during the time window of eEDM
measurement (the last 30\% of the half cycle), so as not to dominate over
the magnetic flux generated by the induced magnetization. 

\begin{figure}[t]
\centering
\includegraphics[width=3.5in, height=2.5in]{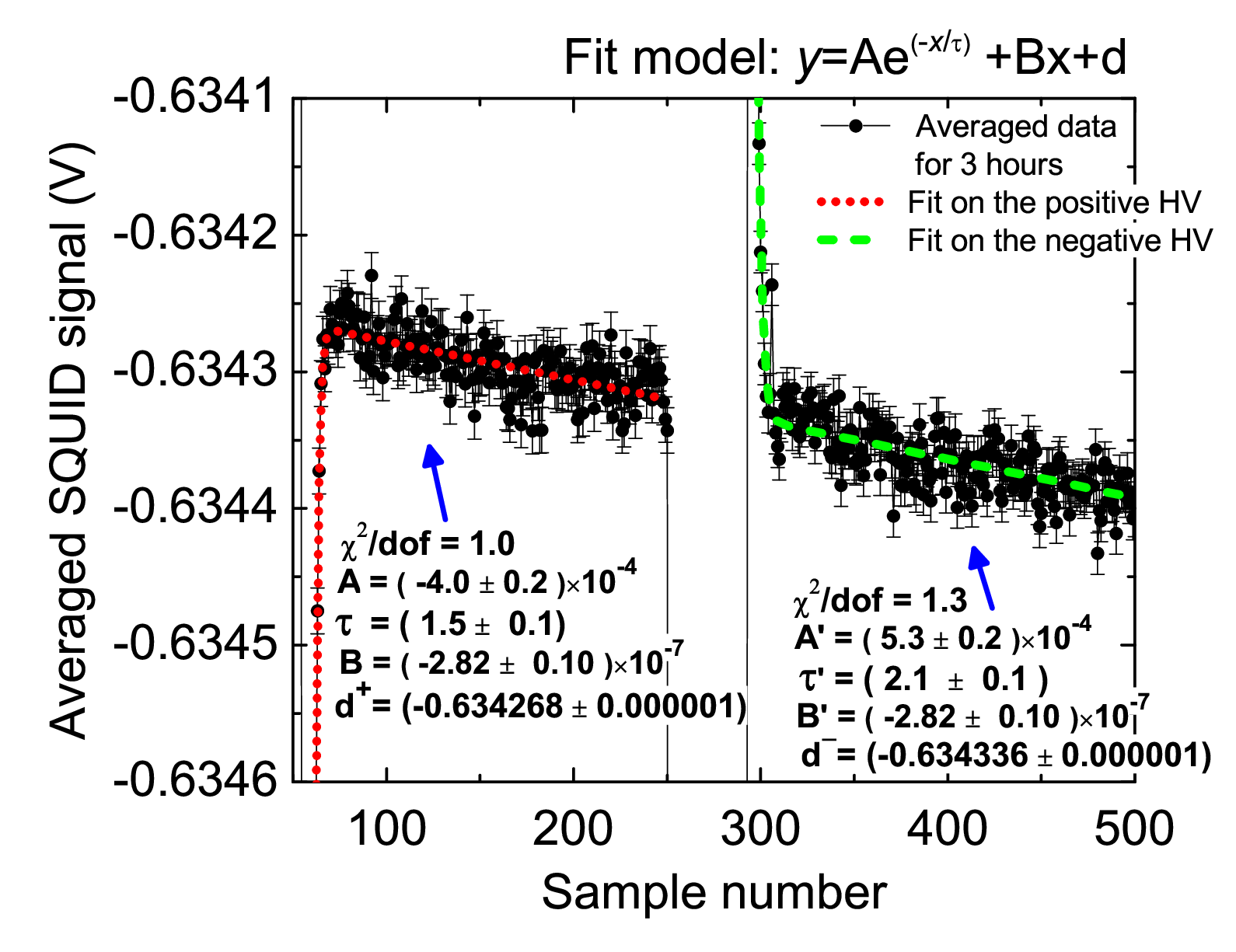}
\caption{\label{fit} SQUID readout (averaged over one non-interrupted run) folded modulo one cycle of field reversal. This data set illustrates the worst case of DC voltage drift. }
\end{figure}

Despite all the improvements discussed above, there still exist
residual voltage drifts. The worst case of such drift is illustrated
in Fig.~\ref{fit}. The DC drift could come from many sources,
including the SQUID electronics, the slow reduction in the level of
the liquid helium, and the pressure drift inside the cryostat, to name
a few.  
Unlike the drift in the HV source, which changes sign in phase with
the HV polarity, this DC drift does not have the same correlation with
the HV cycle.  
To remove this DC drift from the data, we use two independent
algorithms to analyze the eEDM data: drift-correction and fitting.  
The drift-correction algorithm takes the algebraic sum of the SQUID
readout in two adjacent cycles of field reversal, and applies a [+1 -3
+3 -1] weighting to the averaged data for each half cycle. With this,
the effect of the DC drift can be expanded in a polynomial function of
the time, and canceled up to second-order. 
The transient regions at the field reversal are excluded in the data
average. Furthermore, to ensure that the transient current has
sufficiently decayed, the data window contains only the last 30\% of
the time trace (Fig.~\ref{signal}).  
Fig.~\ref{fig:histogram} shows the histogram of eEDM observable,
collected over a typical run with three hours of data. The
distribution can be fit to a Gaussian, giving
$\Phi_{sq}=(-1.89\pm9.25)\times10^{-8}\Phi_0$. 
This corresponds to an eEDM measurement of
$(-0.24\pm1.17)\times10^{-23}$~e$\cdot$cm, after taking into account
all the suppression factors discussed in the previous section.  

The fitting algorithm attempts to fit the entire time trace of the
SQUID readout modulo one cycle of field reversal. Data from repeated
cycles over three hours are averaged to form the overall time-trace
(Fig.~\ref{fit}), and then fit to the following voltage model: 
\begin{equation}\label{fitting}
\begin{array}{l}
V^+(t)=Ae^{-t/\tau}+Bt+d^+, \\
V^-(t)=-A^{\prime}e^{-t/\tau^{\prime}}+(Bt+B\frac{T}{2})+d^-,
\end{array}
\end{equation}
where $V^+(t)$ and $V^-(t)$ are the SQUID readout during the
half-cycle with either a positive or a negative HV polarity.  
The first term in the model characterizes the decay of the transient
current with a time constant $\tau$ or $\tau^{\prime}$.  
These two time constants could be different because of the asymmetry
of the circuit handling the positive and negative voltage, the
difference between the two HV channels, and/or the two HV electrodes.  
The second term describes the DC voltage drift. It is $Bt$ for the
first half cycle, and $Bt+B(T/2)$ for the second half a cycle, which
starts $(T/2)$ later, where $T$ is the period of the field reversal
cycle.  
The final, constant term represents the EDM-induced magnetization and a
DC offset. Note that the sign of the EDM-induced magnetization changes
as the electric field is reversed, while the DC offset remains
constant.  
The eEDM observable is derived simply by taking the difference of the
fitted parameters $d^+$ and $d^-$, 
\begin{equation}
\Delta V= d^+ - d^-.
\end{equation}
The results of the fitting algorithm give $\Phi_{sq}
=(3.07\pm6.34)\times10^{-8}\Phi_0$, corresponding to a $d_e$ of
$(0.39\pm0.81)\times10^{-23}$~e$\cdot$cm. The fitting algorithm
arrives at a better statistical sensitivity than the drift-correction
algorithm because the fitting algorithm uses about 75\% of the
collected data points, as opposed to 30\% used in the drift-correction
algorithm.  
We collected data over two weeks, with a total integration time of
five days.  

\begin{figure}[t]
\centering
\includegraphics[width=3.5in]{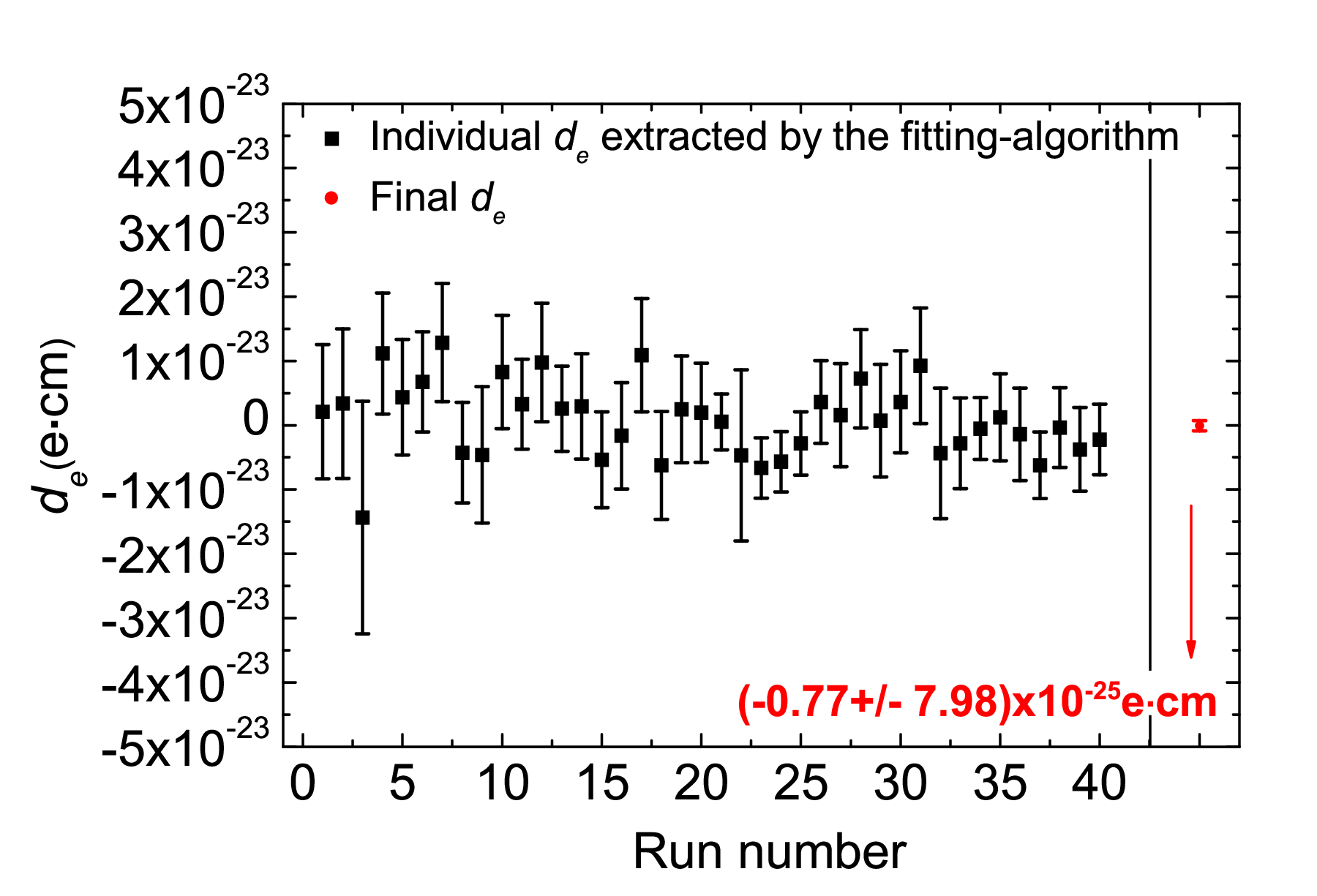}
\caption{ \label{edm} Results of eEDM measurements, processed using the fitting algorithm. Each data point is the averaged result of one run. The final data point is the final average results over 40 experimental runs.} 
\end{figure}

Analysis using the two algorithms shows similar distributions of the
extracted eEDM, and the results using the fitting algorithm are shown
in Fig.~\ref{edm}. Each data point corresponds to a non-interrupted
run lasting for three to four hours. Proper error-weighting is applied
to sum the results of each run to arrive at a final eEDM of
$(0.41\pm1.38)\times10^{-24}$~e$\cdot$cm and $(-0.77\pm
7.98)\times10^{-25}$~e$\cdot$cm, for the drift-correction and fitting
algorithms, respectively.  These results can be compared to the
previous experimental limit using a complimentary solid-state method
in the gadolinium iron garnet system~\cite{Hunter05}. 

\section{Discussion}\label{sys}
\begin{table}
\caption{\label{listsystematic} Systematic effects} % title of Table
\centering % used for centering
    \begin{tabular}{l|c} \hline\hline
    {\bf Source} & {\bf $\Phi_{sq}$ } \\\hline
    Leakage current & $(3.18 \pm 0.07) \times10^{-9}\Phi_0$ \footnote{Average over all runs.} \\
    Displacement current & $ 1.65\times10^{-10}\Phi_0$ \footnote{$C\frac{dV}{dt}<0.2$~pA at 9.4~kV$_{pp}$.} \\
    Remnant magnetization & $<6.02\times10^{-9}\Phi_0$ \footnote{Direct MPMS measurements are limited by the remnant fields.} \\
    Channel cross-talks & $(0.52 \pm 1.51)\times10^{-10}\Phi_0$ \footnote{Channel isolation$>$ 191~dB.} \\
    Vibrational peak @ 2.13~Hz & $<6.02\times10^{-9}\Phi_0$ \footnote{No observable peak on the SQUID noise spectra.} \\
\hline\hline
    {\bf Current Flux Measurement} & $(0.48 \pm 6.02)\times10^{-9}\Phi_0$ \footnote{Averaged over all runs.} \\ \hline\hline

\end{tabular}
\end{table}

\begin{figure}[t]
\centering
\includegraphics[ width=3.4in]{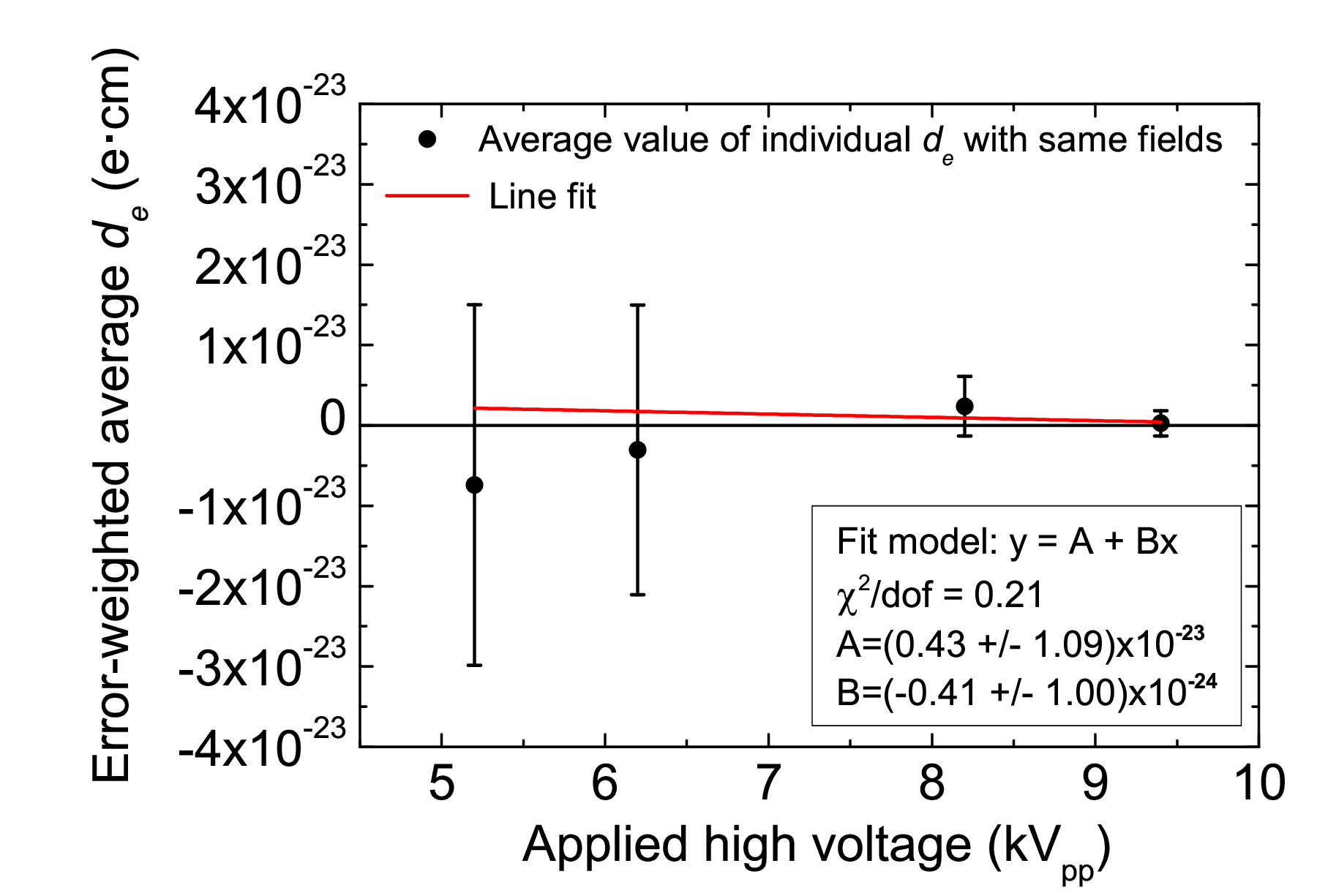}
\caption{ \label{systematic} Measured eEDM as a function of the applied HV. The data point at 9.4~kV$_{pp}$ contains the highest statistics.}
\end{figure}

A comprehensive list of systematic effects is shown in
Table~\ref{listsystematic}. Since the measured physical observable is
the magnetic flux, we compare the spurious flux generated by each
known systematic effect. The dominant effect is the leakage current
through the sample which produces a magnetic field in phase with the
polarity of the HV. 
To first order, the magnetic field generated by the leakage current is
perpendicular to the EDM-induced magnetization, some fraction of the
field, however, can be measured by the SQUID sensor due to the slight
tilt of the pickup coil. Surface currents forming a helical path would
generate additional magnetic flux. Studies of the correlation between
the displacement currents (during field transients), the
time-derivatives of the applied HV, and the SQUID signal allow us to
separate the contributions from the displacement current from that of
the leakage current (due to the finite resistivity of the
sample). These studies show that about 1.4\% of the radial field
generated by the the leakage current could leak into the pickup coil
and contribute to a spurious signal. For example, a leakage current of
$(6\pm2)$~pA (averaged over three hours) measured at the maximum
applied voltage of 10~kV$_{pp}$ generates a spurious magnetic flux
$\Phi_{sq}$ of $(3.79\pm1.26)\times10^{-9}\Phi_0$. Despite the finite
leakage current, the measured eEDM is shown to be independent of the
strength of the applied electric fields within the error bars, as
shown in Fig.~\ref{systematic}. Most measurements were made at
9.4~kV$_{pp}$. This suggests that the experiment is free of systematic
effects linear in the HV.  

We also attempted to characterize the magnetic hysteresis of the GGG
sample using the SQUID-based susceptometer system,
however, due to the 
large variations of the remnant field upon each field ramping, these
measurements cannot be used to determine the magnetic hysteresis of
the GGG sample to the sensitivity level required in our experiment.  
Since there is no measurable offset between the half cycles, we can
place an upper limit on the remnant magnetization using our EDM
results.  
Given the knowledge of the magnetic susceptibility of GGG, we need to
control the remnant magnetization, which changes in phase with the
applied field, to below $2.5\times 10^5\mu_{B}$/cm$^3$. 

The dominant systematic effect is the leakage current, which creates
an additional magnetic flux, and can lead to a spurious eEDM signal
through Eq.~\ref{phi}. 
The measured flux for each run is corrected by subtracting the
additional magnetic flux created by the leakage currents, before
extracting the eEDM value. 
The total systematic effect is estimated to be $(4.80\pm
0.12)\times10^{-25}$~e$\cdot$cm with the proper error-weighting.  
This leads to the final reported eEDM value of
$(-5.57\pm7.98_{stat}\pm0.12_{syst})\times10^{-25}$~e$\cdot$cm with
five days of data averaging, in this prototype experiment running at
4.2~K. 

\section{Conclusions and Future Work}
In this work, we have learned to solve several systematic problems, and
demonstrated the feasibility of the solid-state method using the
paramagnetic insulator GGG at 4.2~K. We report our first background-free
experimental limit on the eEDM. Further improvement in the sensitivity
should certainly be possible.  We consider each term in Eq.~\ref{phi} for the
eEDM-induced flux.

We have developed a second-generation HV system
capable of generating 40~kV$_{pp}$ (Sec.~\ref{sec:improvements}).
This will increase the applied electric field $E_{ext}$ by a factor
of 4. 

One factor in the geometric flux suppression term $f$ is the coupling
efficiency of the pickup coil to the sample.  This can be improved to
near unity by winding the pickup coil around the midplane of the
sample. 

The sensitivity to $\Phi_{e}$ is limited by the SQUID
intrinsic noise.  We can expect this to improve by about a factor of 2
(to 3~$\mu \Phi_{0}/\sqrt{Hz}$), using a state-of-the art sensor.  

More ambitiously, the engineering design of the second-generation
experiment includes 10 sample/electrode modules very similar to those
in the experiment reported here.  Assuming a parallel configuration in which
the same SQUID 
coupling efficiency ($\beta$ in Eq.~\ref{squid}) can be attained, this
will increase the 
effective area $A$ in Eq.~\ref{phi} by an order of magnitude without
otherwise compromising the flux sensitivity.  More
tentatively, it 
should be possible to optimize the input and SQUID loop inductances to
improve $\beta$ by about a factor of 10.

Taken together, these improvements imply a projected sensitivity of
the experiment within an order of magnitude of the molecular
result~\cite{Acme14}. 
This assumes no change in the sample susceptibility. As detailed
in~\cite{Schiffer95}, the Curie-Weiss behavior of the susceptibility
as 
observed in Fig.~\ref{1} can be expected to continue down to temperatures of
about 1~K, implying an improvement in this quantity by at least a factor of
2.  A more rapid increase can be expected at lower temperatures
on account of a spin-glass transition, with an observed maximum of
about 0.15 (cgs units) near 0.2~K for single-crystal samples.  We note, however,
that the demagnetization 
factor (Eq.~\ref{chi}) is nearly saturated in our chosen sample geometry,
making improvements in susceptibility of little value without
significant alteration of our practical design.   

A more promising route to greater sensitivity is afforded by reducing
the sample thickness, and accepting a modest increase in the
demagnetization factor while realizing a linear improvement the
electric field strength.  Reducing the sample thickness by a factor of
2 or more would, with the additional changes above, yield a
sensitivity greater than the molecular limit. 
This assumes, of course, that systematics associated with leakage
currents can still be controlled.  Operation at 1~K would
increase the resistivity of the sample and could well be of additional
benefit for this approach, though the systematic effects will warrant
careful study.  

\section{Acknowledgments}
We would like to thank K. McClellan for assisting with the GGG sample
growth and preparation, and J. Valdez for performing X-ray diffraction
measurements on the GGG samples.  This work is supported by NSF grants
PHY-0457219, PHY-0758018, and the Indiana University Center for
Spacetime Symmetries (IUCSS).

\bibliography{PRD_ver5jcl_2}
%\bibliography{eEDM}

\end{document}